# Applying Normalization Process Theory to Explain Large-Scale Agile Transformations

**Noel Carroll**
Lero, National University of Ireland Galway,
Galway, Ireland
noel.carroll@nuigalway.ie

**Kieran Conboy**
Lero, National University of Ireland Galway,
Galway, Ireland
kieran.conboy@nuigalway.ie

**ABSTRACT**

Given the prevalence and effectiveness of agile methods at a team level, large organizations are now attempting to mimic this success at large-scale by adopting large-scale methods such as Scaled Agile Framework (SAFe), Spotify, and Large-Scale Scrum (LeSS). However, compared to insights on traditionally small-scale methods, the extant literature provides sparse coverage on theories to examine large-scale agile transformations. In this article, we focus on the challenge of normalizing large-scale agile transformations and apply Normalization Process Theory (NPT) to support theorize about this process. We present our initial case study findings and outline future research on the application of NPT for large-scale transformations. From a research and practice perspective, we explain how NPT can be adopted to focus on the processes of embedding and sustaining practices – activities which are very often ignored, yet central to the success or failure of transformations.

**LARGE-SCALE AGILE TRANSFORMATION**

Agile methods have been well received by practitioners and academics over the past decade. The benefits of agile methods were originally believed to best suit in their traditional form, i.e. in small and collaborative single-team projects with evolving needs (Abrahamsson et al. 2009; Boehm and Turner, 2005; Dybå and Dingsøyr, 2008; Wang et al. 2012). However, given the success of traditionally small-agile methods, many large software organizations have begun to adopt the methods in a large-scale context (Laanti et al. 2011). In reality, this poses significant challenges considering teams are often fragmented across a large organization (Dikert et al. 2016) and adopt many variants of agile methods (Conboy and Carroll, 2019). In addition, the adoption of agile methods can emerge from top-down (management-driven) or bottom-up (team-driven) and various factors can obstruct clarity regarding the reasons for initiating large-scale agile transformations (Dikert et al. 2016). As large organizations face growing pressures and expectations to scale, there is a natural tendency to ensure teams are more coherent by scaling agile methods to leverage cost savings and dynamism at scale. As a result, large-scale agile methods such as Scaled Agile Framework (SAFe) (Leffingwell, 2019) and Large Scale Scrum (LeSS) (Larman and Vodde, 2019) are increasingly prevalent in contemporary software organizations (Dikert et al. 2016; Dingsøyr and Moe, 2014; Kalenda et al. 2018; Lagerberg et al. 2013; Paasivaara et al. 2018). Such methods in large-scale environments present many challenges as complex socio-technical interdependencies increase with the organizations size (Dybå and Dingsøyr, 2008; Livermore, 2008; Misra et al. 2010; Rolland et al. 2016). While there are many potential benefits, large-scale agile transformations are fraught with challenges. For example, Conboy and Carroll (2019) identify nine key challenges associated with implementing methods on a large-scale, including: difficulty in "comparing and contrasting large-scale methods"; lack of "readiness and appetite for large-scale methods"; the need to "balance organizational structure while adhering to large-scale methods"; and, the lack of "evidence-based use of large-scale methods". Therefore, we are at risk of repeating the same implementation and adherence issues as that experienced within a small-scale context (Conboy and Carroll, 2019; Dikert et al. 2016; Lindvall et al. 2004; Rolland et al. 2016). Thus, the objectives of this study are:

   (i) *To identify what assumptions exist around large-scale agile transformations in information systems development*
   (ii) *To explain what factors across information systems development organizations enable or inhibit the normalization of a large-scale agile transformation*

**RESEARCH MOTIVATION**

Focusing on normalization is important and novel as it allows us to determine "*the work that actors do as they engage with some ensemble of activities and by which means it becomes routinely embedded in the matrices of* *already existing, socially patterned, knowledge and practices*" (May and Finch, 2009; p. 540). This is important because empirical evidence regarding the adoption of methods for large-scale transformations, their use, effectiveness, and challenges is still very much in its infancy. Specifically, the motivation to study normalization emerged from literature and the need to:
1. Re-examine assumptions of large-scale agile transformations
2. Emphasize sustainability over adherence of large-scale agile transformations
3. Unpack the dynamic nature of large-scale agile transformations
4. Build a cumulative tradition on information systems (IS) theories

This research sets out to empirically examine the fundamental assumptions around the normalization of large-scale agile methods with a view to develop a theoretical foundation to better explain how methods are best sustained in practice. Much of the literature on large-scale transformations emphasize the role of method adoption, and outlines various benefits and challenges associated with adherence to method variations. However, the use of agile methods in large development projects remains a significant challenge with little empirical evidence of successful cases (Cao et al. 2004; Conboy and Carroll, 2019; Dikert et al. 2016). Yet, we have become accustomed to many assumptions underlying current research on large-scale agile development which calls for the need to reconceptualize what comprises of (i) large-scale agile methods (Rolland et al. 2016) and (ii) the large-scale transformation process. However, little if any research efforts have focused on the normalization phenomenon of large-scale transformations. To expand on this rationale, we consider the various shortcomings across research on large-scale agile transformations.

**Normalization Re-examines Assumptions of Large-Scale Agile Transformations**. Normalization explicitly examines weak underlying assumptions of embedding change such as large-scale agile transformations. Yet, there is a lack of guidance on how to find the optimal degree of transformation. In order to better understand large-scale agile development, Rolland et al. (2016) explain that we need to challenge the assumptions that underpin large-scale projects as they pertain to the adoption of agile methods. This is of particular importance within a large-scale agile transformation context as agile methods emerged largely from practice without a great deal of consideration for sound conceptual foundation for the topic (Abrahamsson et al. 2009; Ågerfalk et al. 2009; Conboy 2009; Conboy and Carroll, 2019; Schnitter and Mackert, 2010). Normalization proposes a working model of implementation, embedding and integration in conditions marked by complexity and emergence (May and Finch, 2009). This is important given the lack of research regarding incorrect assumptions on large-scale agile transformations (Rolland et al. 2016).

**Normalization Emphasizes Sustainability over Adherence of Agile Methods**. Formal large-scale agile methods, such as SAFe, Scrum at Scale, or Spotify, creates a tendency to measure transformation by adherence to a specific method, rather than the value it provides to sustaining practice (Conboy and Carroll, 2019; Dikert et al. 2016). Normalization allows us to focus on the specific work (actions, processes or events) of embedding and of sustaining practices through specific agile methods which helps in understanding why some methods become normalized while others do not (May et al. 2009). Normalization invites us to move beyond adherence and address incorrect assumptions by examining the fine-grained approaches to sustaining methods as opposed to one-off method use.

**Normalization Unpacks the Dynamic Nature of Large-Scale Agile Transformation**. The nature of agile principles (Beck et al. 2001) were developed in response to guide organizations on how to operate in small team and highly dynamic environments. Yet, research on large-scale agile transformation fails to consider the dynamic nature of large-scale agile methods (Dingsøyr et al. 2018) which comprise of, for example, larger teams, rigid organizational structures, and regulations. Nowadays, businesses operate in ever-changing dynamic environments making it challenging to normalize transformations – almost as if chasing a 'moving target' (Posen and Levinthal, 2012). Thus, there continues to be a widening research gap on the need to normalize agile methods at scale while continuously reacting to a dynamic and influential business environment. Normalization, specifically NPT, offers an alternative yet complementary research lens to existing IS theories as an overarching theory using 16 sub-components. Therefore examining the normalization of large-scale agile transformations is not a case of "*old wine in new bottles*" but rather, given the growing and relentless pace and scale of change across global software organizations, it calls for new insights on how transformations are implemented, embedded, integrated, and evaluated within software organizations for continuously shifting business and software environment (Fitzgerald and Stol, 2017; Leffingwell, 2007).

**Normalization Builds a Cumulative Tradition on IS Theories**. Historically, the IS discipline has been weak in developing a cumulative research tradition (Benbasat and Zmud, 1999; Keen 1980). This is also true within an agile context (Abrahamsson et al. 2009; Conboy 2009) which makes it challenging to develop and assess strong theoretical models that translate into practice. For example, agile development methods have been described as examples of success stories that seem to have "*run counter to the prevailing wisdom*" in information systems development (Ågerfalk et al. 2009; p. 317). In practice, software organizations typically select from a range of large-scale methods and custom fit its applicability to facilitate transformations and report on varying benefits and drawbacks of selected methods. Yet there is clearly a lack of a cumulative tradition within the IS research community to communicate insights on the scaling of transformations (Conboy and Carroll, 2019) and potentially reopening what was once described as a 'credibility gap' in IS research relevance (Benbasat and Zmud, 1999). Current thinking on large-scale transformations fails to explain the importance of embedding, sustaining and normalizing large-scale transformation methods and poses questions as to whether the relevancy of research implications are implementable.

## NORMALIZATION PROCESS THEORY

### NPT Theoretical Constructs

NPT is a derivative sociological theory on the implementation, embedding and integration of new technologies and organizational innovations (May and Finch, 2009). NPT identifies factors that promote and inhibit the routine incorporation of complex interventions into everyday practice (May et al. 2009; Murray et al. 2010). It focuses on providing practical insights on specific phenomena both qualitatively and quantitatively. NPT also explains how transformations operate, seeking not only at early implementation, but beyond this to the point where an intervention becomes so embedded into routine practice that it 'disappears' from view (i.e. it is normalized) (Murray et al. 2010). Specifically, NPT is concerned with the social organization of the work (implementation), of making practices routine elements of everyday life (embedding), and of sustaining embedded practices in their social context (integration) (May 2006). NPT allows us to examine assumptions and dynamics of large-scale agile transformations with a view to sustain agile methods in practice and build a cumulative tradition on IS theories.

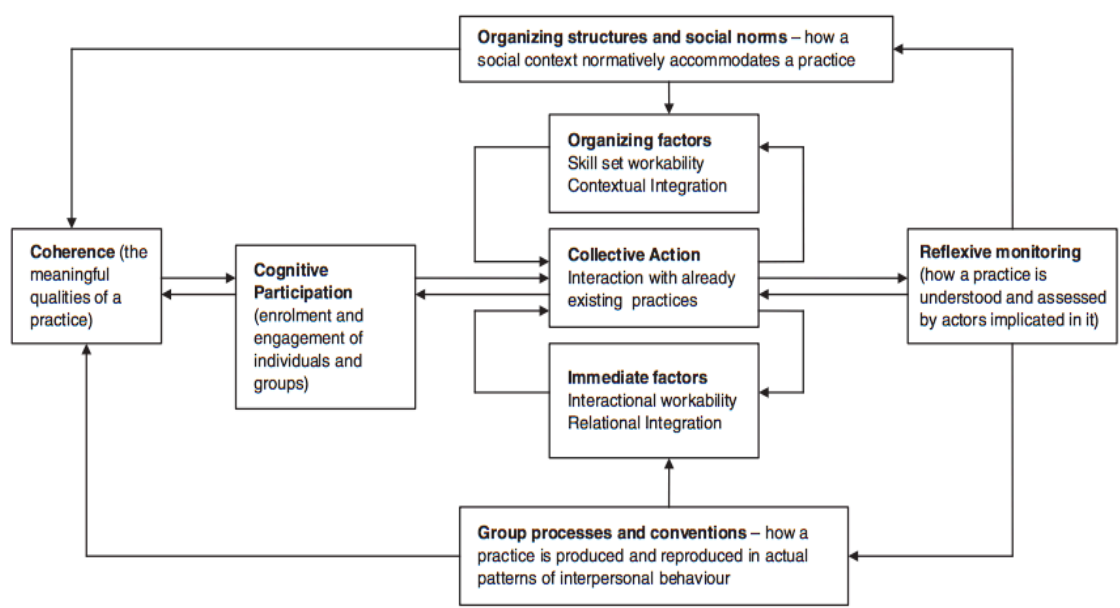

**Figure 1. Model of Normalization Process Theory components (May and Finch, 2009)**

NPT can support how the IS community theorize about transformations (Figure 1). NPT has clear applicability to large-scale agile transformations to examine the adoption of a new model of practice through the following theoretical constructs:

1. **Coherence**: This construct refers to the process of sensemaking that individuals and organizations undergo in order to promote or inhibit the routine embedding of a practice (i.e. determining specifically "*what is the work?*"). This allows us to examine the rationale and drivers to transform an organization, i.e. how people understand and make sense of a practice and its implications on defining and (re)organizing a practice.
2. **Cognitive Participation**: This construct examines how stakeholders engage in their work in order to enroll individuals in the newly adopted practice (i.e. "*who does the work?*"). This allows us to identify the roles and responsibilities which are developed to sustain and participate in transformations.
3. **Collective Action**: This construct focuses on the work that individuals and organizations have to do to change practice by enacting the new practice (i.e. "*how does the work get done?*"). This allows us to examine the specific practices, organizing factors, and tools used enact and sustain new practices within a transformation working towards the same vision.
4. **Reflexive Monitoring**: This construct describes the value realization inherent in the informal and formal appraisal of a new practice and the reported process improvements resulting in the transformation process (i.e. "*how is the work understood?*"). It assesses its advantages and disadvantages and develops stakeholders' comprehension regarding the value and overall effect of the new practice. This allows us to determine which evaluation methods are used to guide how executives reflect or appraise the effects of a transformation. This can also shed new insights on organizing structures, social norms, group processes and conventions, i.e. assessing patterns of work and outcomes.

| Core Construct of NPT | Construct Components of NPT |
|---|---|
| **1. Coherence:** sensemaking individually and/or collectively when faced with the problem of operationalizing a set of practices. | **1.1 Differentiation**: This is an important element of sensemaking to understand how a set of practices and their objects are different from each other. |
| | **1.2 Communal specification**: Sensemaking relies on people working together to build a shared understanding of the vision, aims, objectives, and expected benefits of a set of practices. |
| | **1.3 Individual specification**: Sensemaking has an individual component whereby team members need to understand their specific tasks and responsibilities around a set of practices. |
| | **1.4 Internalization**: Sensemaking involves organizational stakeholders understanding the value, benefits, and importance of a new set of practices. |
| **2. Cognitive Participation:** the relational work that people do to build and sustain a community of practice around a new technology or method. | **2.1 Initiation:** When a set of practices are new or modified, a core challenge is to examine whether key team members are contributing to drive the transformation process forward. |
| | **2.2 Enrolment:** Team members may need to organize or reorganize themselves and others in order to collectively contribute to the work involved in new practices. |
| | **2.3 Legitimation:** An important component of relational work around participation is the task of ensuring that other team members believe it is right for them to be involved, and that they can make a valid contribution to the transformation process. |
| | **2.4 Activation:** Once a transformation is underway, team members need to collectively define the actions and procedures needed to sustain the new practice and to stay committed to the vision of the transformation. |
| **3. Collective Action:** the operational work that people do to enact a set of practices (e.g. adhering to a new method). | **3.1 Interactional Workability**: This refers to the interactional work that team members do with each other, with artefacts, with other elements from a set of practices and seek to operationalize them in everyday settings. |
| | **3.2 Relational Integration**: This refers to the knowledge that people generate to build accountability and maintain confidence in a set of practices and in each other as they use them. |
| | **3.3 Skillset Workability**: This refers to the allocation of tasks that underpins the division of labor typically built up around a set of practices as they are operationalized in the real world. |
| | **3.4 Contextual Integration**: This refers to the resource work, i.e. managing a set of practices through the allocation of different kinds of resources and the execution of protocols, policies and procedures. |
| **4. Reflexive Monitoring:** the appraisal of work that people do to assess and understand the ways that a new set of practices affect them and others around them. | **4.1 Systematization**: Team members in any set of practices may seek to determine how effective and useful it is for them and for others. This involves collecting information in a variety of ways. |
| | **4.2 Communal appraisal**: Team members work together (formal or informal collaborations) to evaluate the value of a set of practices through experiential and systematized approaches. |
| | **4.3 Individual appraisal**: Team members working within a new set of practices also work experientially as individuals to appraise its effects on them and the contexts in which they are set. |
| | **4.4 Reconfiguration**: Appraisal work by individuals or groups may lead to attempts to redefine procedures or modify practices. |

**Table 1. Applying NPT to Large-Scale Agile Transformations (adapted from May and Finch, 2009; May et al. 2009)**

NPT provides a rich theoretical lens to explain a transformation process since it allows us to uncover how practices become routinely embedded in their social contexts as the result of people working, individually and collectively, to enact them (May and Finch, 2009; May et al. 2009). Within each of the core theoretical constructs, we can examine the normalization of large-scale agile transformations and sheds new insights on organizing structures, social norms, group processes and conventions, i.e. work relating to assessing patterns of work and outcomes. We argue that NPT can support the IS community to theorize about the transformation process which encapsulate the four theoretical constructs. Within each of the NPT core theoretical constructs, there are four additional components (16 in total) which can support to explain large-scale agile transformations (see Table 1).

## METHODOLOGY

### Research Design

Adopting a qualitative research approach provides an opportunity to utilize different qualitative research designs to obtain key insights from the potentially broad scope which normalization comprises of. Given the under-researched nature of normalization in large-scale agile transformations, we employed a single case study approach (Yin, 1984). A case study examines a phenomenon in its natural setting, employing multiple methods of data collection to gather information from one or a few entities (people, groups, or organizations) (Benbasat et al. 1987). For the purposes of this study, we adopt a single-case study as a revelatory and unique case (Yin, 1984) for large-scale agile transformation. The case study focuses on a global financial service organization, FinanceCo (pseudonym to protect anonymity) which comprises of approximately 1,250 employees across the U.S, Ireland, India, and China. Over the past few years, FinanceCo have experimented with various agile methods and customized agile methods. In the past two years, FinanceCo underwent a large-scale agile transformation strategy and adopted Spotify. This exploratory single case study focused on the Spotify method as the unit of analysis and uncovers individuals experience during the large-scale transformation process.

### Data Collection

Within FinanceCo, AgileTeam (pseudonym team name) were assigned as the field researcher's main point of contact. The researchers were provided with access to staff across the team and digital records (historical, real-time resources, and strategic plans). The researchers were introduced to their team members (see examples of interviewees profiles in Table 2); agile project management systems (e.g. JIRA); team meetings and other agile-related events. Given the level of such access to AgileTeam, we could obtain several sources of evidence through documentation, archival records, interviews, and direct observation to verify interview accounts.

| Code | Position | Years Experience | Code | Position | Years Experience |
|------|----------|------------------|------|----------|------------------|
| SVP | Senior Vice President | 22 | SSE1 | Senior Software Engineer | 15 |
| STM | Senior Technical Manager | 13 | SSE2 | Senior Software Engineer | 9 |
| DAM | Data Analyst Manager | 13 | SSE3 | Senior Software Engineer | 8 |
| TM | Technology Manager | 15 | SM | Scrum Master | 6 |
| PM | Project Manager | 8 | JSE1 | Junior Software Engineer | 3 |
| MS | Manager at Spotify | 3 | JSE2 | Junior Software Engineer | 2 |

**Table 2. Summary of Interviewee Profiles**

For this research, we piloted 12 semi-structured interviews (11 across AgileTeam and one from Spotify) which broadly encapsulates the large-scale agile transformation process. There are several strengths in interviewing and for the purposes of this exploratory study, it permits the respondent to move back and forth on the interviewee's experiences and perceptions (Glaser & Strauss, 1967). The flexibility of the technique allows the researcher to probe, to clarify, and to create new questions based on what has already been heard.

### Data Analysis

Given the NPT framework presented in Figure 1, four key constructs which are of importance in our initial data analysis are (i) coherence; (ii) cognitive participation; (iii) collective action; and (iv) reflective monitoring. We reviewed the interview transcripts and categorized key insights into the four NPT constructs. Analysis was necessary from the first interview because it was used to support and direct the next interview and observations.

NPT also provided a diagnostic frame to categorize various benefits, challenges and recommendations when interviewees articulated their experiences with the large-scale agile transformation. As illustrated in Figure 2, we employed open coding to identify the relationships among interview statements and observations with NPT construct categories. We then employed axial coding to examine how NPT component categories were related to their subcategories and their relationships before moving to selective coding to theorize on large-scale agile transformations.

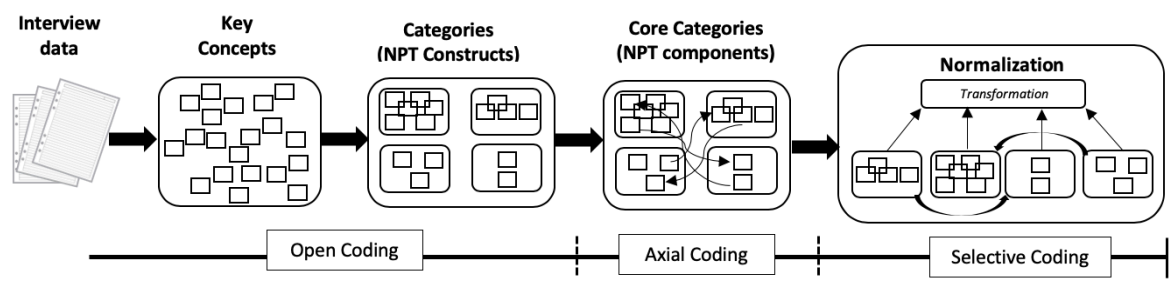

**Figure 2. Levels of abstraction to theorize on large-scale agile transformation**

**PRELIMINARY RESULTS**

This section presents four indicative findings around adhering to large-scale agile methods versus adherence and presented within an NPT lens (under the four core theoretical constructs and 16 components previously listed in Table 1). In this case study, our preliminary findings uncover key factors which prevented the normalization of a large-scale agile transformation and eventually led to the abandonment of the agile method.

**Coherence in a Large-Scale Agile Transformation**

Our findings indicate that there was a lack of coherence throughout the AgileTeam at FinanceCo regarding the justification for (a) transforming the organizations, and (b) the adoption of Spotify as a large-scale agile method.

*Differentiation:* FinanceCo collectively engrained a loose sensemaking approach when faced with the problem of operationalizing a new agile method, i.e. Spotify. For example, a Data Analyst Manager (DAM) referred to the lack of coherence around the differentiation of Spotify from previous customized agile methods, yet was tasked with developing metrics around the new agile method:
> "I'm not entirely sure what this method will achieve or how it will be realistically implemented for large-scale agile. For me, it's difficult to identify what makes Spotify different to how we used to work here. I guess it's a little bit of trial-and-error, but I need to define performance indicators on what we should expect from Spotify for this large-scale transformation?"

*Communal specification:* Software engineers followed similar sentiments around communal specification on their coherence to participate in the transformation process. For example, one Senior Software Engineer (SSE2) explained that:
> "I think it should be a case of working out what went wrong before [at small-scale], as I felt it was going well? I don't get any real insight as to what's driving the need for this change to a large-scale method, or how this changes what we do."

*Individual specification:* On several instances interviewees expressed a need to be more involved in the decision-making process around the rationale for transformation to accommodate for individual specification. However, the Senior Vice President (SVP) pointed out that the rationale for transformation goes beyond team-level and is a decision primarily concerned at an executive level:
> "We often talk about being agile but within a large-scale context, how agile are we organization-wide? This is an executive decision. When we explored the 'Spotify Model', we saw how we could scale agile beyond the team-level to the wider organization which brings in new principles and practices."

*Internalization:* There were also contrasting interpretations as to how team members internalize the value of adopting Spotify and its impact on their working environments. For example, one Senior Software Engineer (SSE1) explained that:
> "I learned about the large-scale transformation through a PowerPoint presentation at a meeting. I have been working here for the last 5 years and this new model [Spotify] means I have to change how I

*work…and where I now sit in the room (in various Tribes and Squads). Realistically, how does this change how I will develop software?"*

In contrast, a Senior Technical Manager (STM) explained that they saw potential in the large-scale agile model:
*"We can now operate under a cool model where we are divided into Squads, Tribes, Chapters and Guilds. Across the company, this should also promote a greater sense of autonomy, ownership and accountability across teams. That should be good for everyone."*

Based on these initial interviewee insights, it became apparent that at the very early stages of the FinanceCo large-scale agile transformation, there was a lack of coherence among the AgileTeam regarding the justification for (a) transforming the organizations, and (b) the adoption of Spotify as a large-scale agile method.

**Cognitive Participation within a Large-Scale Agile Transformation**

The findings indicate how FinanceCo largely focused on adopting the Spotify method as a structured approach to guide the implementation of new team structures (e.g. Squads, Tribes, Chapters, and Guilds), rules and processes to encourage small and frequent releases.

*Initiation:* One Junior Software Engineer (JSE1) described their experience from the initiation of Spotify in the large-scale transformation process and explained that:
*"On one hand, we are told that as a Squad, we are expected to self-organize, and determine the best way to work, for example, from Scrum Sprints to Kanban…and then on the other hand after investing significantly in new tools we are also expected to tailor a hybrid approach around project management tools…and now new metrics for the large-scale transformation?"*

*Enrolment:* A Junior Software Engineer (JSE2) outlined their concerns around the lack of opportunity to have any input on how they are enrolled into the newly transformed agile work practices:
*"I know that agile promotes the need to be collaborative but not all collaboration is productive especially in large-scale. I'm frustrated with all the [global] meetings, ceremonies, and checkbox exercises we have to endure which, I can see effects how I now engage with colleagues."*

*Legitimation:* Championing the transformation process proved to be a key factor to ensure buy-in across teams. To do so, it was critical to demonstrate how Spotify presented a legitimate large-scale agile transformation for FinanceCo. For example, the Senior Vice President (SVP) described how:
*"We need to continuously evolve. Having identified the potential for Spotify to scale here, we need to sell it an executive level and demonstrate how it not only meets our needs locally, but how other global sites and business units should follow-suit to transform at a large-scale."*

*Activation:* During the large-scale agile transformation, there were significant efforts by team members to define the actions and procedures needed to ensure adherence to the Spotify method. For example, one Senior Technical Manager (STM) reassured that:
*"Since we [FinanceCo] had increased capabilities to capture evidence on global software productivity, this became a core driver for 'healthy conversations' such as redefining our definition of 'Done' and keeping the transformation alive across the organization."*

However, it was extremely challenging for global team members to explain how their commitments would essentially sustain the large-scale agile transformation (using the Spotify method). For example, one Project Manager (PM) described:
*"We regularly hear from senior management that it is vital to keep on track for the transformation process. But I don't really know what the endgame is or what success looks like? I'm frustrated trying to juggle and constantly reprioritize tasks – and we don't have a roadmap. If I can't connect the dots, how will my team members do so (locally and globally)?"*

The researchers also had the opportunity to discuss some of the challenges experienced with AgileTeam with a Manager at Spotify (MS). He explained that:
*"I must caution that many organizations look at the Spotify model as being their panacea which will solve all of their problems – and use it 'out of the box'. The Spotify model worked for Spotify! It was not designed to be adopted by every organization. You must carefully consider what elements work for you and which don't."*

We can identify how AgileTeam failed to consider what processes and techniques could be adopted to successfully sustain their large-scale transformation process and potentially over-relied on adhering to the Spotify method.

**Collective Action throughout a Large-Scale Agile Transformation**

Within AgileTeam, adhering to actions prescribed through Spotify encouraged small and frequent releases and decouple releases (through various Squads).

*Interactional Workability:* From a normalization perspective, interactional workability plays a key role to examine the interactional tasks that team members carryout and their specific actions to operationalize them during a large-scale agile transformation. However, Spotify does not promote standardization, but rather the need for cross-pollination of actions to transform practices. For example, the AgileTeam Squads were encouraged to use specific tools such as JIRA to monitor the flow of software productivity and performance and embrace new practices such as converting wall space into whiteboard space. The rationale for doing so meant that the success of such practices met less resistance when other Squads tried to adopt them and eventually became a default standard. However, as one Senior Software Engineer (SSE1) described, for high performing developers there was a dilemma in adhering to methods which hampers their overall performance:
> "I do like some change with the Spotify method, but we have to do so many new and meaningless things that I'm spending less time developing software and the metrics will show that. We meet so often now and we, as a team, have actually slowed down as we try to scale agile."

*Relational Integration*: We also learn from FinanceCo that relational integration was something that was well executed across AgileTeam in the short-term. For example, training was provided at a team-level and the introduction of new metrics instilled a greater sense of knowledge-transfer and accountability to maintain confidence in Spotify as a method for large-scale transformation. However, in the long-term, teams could not appreciate nor evaluate how their continuous improvement efforts were contributing to the overall large-scale transformation process. For example, one Senior Software Engineer (SSE3) explained that:
> "The metrics are telling us we are getting faster, but is faster better? Many of us on the team feel like we're just on a hamster wheel with growing expectations to support the large-scale transformation, focusing on how fast we work and moving products to 'Done'. The metrics are good, but we don't understand how these fit into the Spotify model?"

One explanation for this may be the lack of management training on large-scale agile transformations. For example, at an executive level, the Senior Vice President (SVP) expressed concern about the lack of training to guide the overall transformation process:
> "There are so many opportunities for training on how to operationalize Spotify such as Scrum Masters, one-day training programs, and other events but we don't have any training programs at an executive level to actually manage the large-scale transformation process."

*Skillset Workability:* From a normalization perspective, Spotify did alter the skillset workability (i.e. the allocation of tasks that underpins the division of labor to operationalize the Spotify method). Within FinanceCo the new Spotify team structures such as Squads, encouraged team members to sit together, and share skills and tools throughout the software production process. One Project Manager (PM) explained that:
> "We are wondering how we deal with the notion of being a self-organizing team and decide our own way of working especially at large-scale. Regardless of using Scrum sprints, Kanban, or some mix of these approaches, we need to learn how to effectively utilize these under this [Spotify] model. We are trying to preserve this 'scaled independence' of software teams to self-organize."

Yet, within one of the Squads, there was a growing sense that the Spotify structure removed accountability and responsibility within teams. For example, one Junior Software Engineer (JSE1) described this as:
> "I feel we are losing some sense of accountability as it became the responsibility of a Tribe for certain deliverables. When delays occur, blame can then quickly shift across the cross-functional teams making it really difficult to identify the effectiveness of team members, tools, and decision-making to remove impediments."

*Contextual Integration:* Significant efforts were placed on contextual integration within AgileTeam to adhere to the Spotify method, yet strict adherence to the method had proven to both help or hinder team performance. This largely depended on the team experience level and acceptance among individuals regarding the large-scale transformation process. To address potential issues from impacting on the transformation strategy, the Senior Vice President (SVP) suggested that:
> "Transparency alone initiates improvements. Once [software productivity] data became assessable, visible and presented in a manner in which teams consume and learn how they were performing, they instinctively put actions in place to improve the large-scale transformation process."

**Reflective Monitoring on a Large-Scale Agile Transformation**

Within FinanceCo the appraisal of team performance to better assess and understand the large-scale transformation progress and efforts on continuous improvements across AgileTeam was important.

*Systematization:* Across AgileTeam the systematization of Spotify was important to all team members in order to determine how effective the method was for them and their colleagues. However, much of the focus regarding systematization focused on analytical accounts (such as performance dashboards) but failed to account for the rising negative sentiments towards the large-scale agile method in the early transformation stages. For example, one Junior Software Engineer (JSE2) explained that:
> "Our metrics and key performance indicators inform us that we are getting 'faster' which must be a good story for the transformation process. But I am not happy in my work. Many of my colleagues feel the same and there is a lower morale than that of pre-Spotify."

*Communal appraisal:* AgileTeam attempted to informally research and collect themes from individual anecdotal accounts of positive and negative experiences throughout the large-scale transformation process. From a normalization perspective, communal appraisal is vital as team members work together (formal or informal collaborations) to evaluate the value of a new large-scale agile method. Yet, within AgileTeam, one Senior Software Engineer (SSE1) referred to a growing paradoxical sense that:
> "We are self-organized to a point. We can drive our own performance. But by reaching new performance targets or levels through those new team structures which Spotify impose, I do feel we are victims of our own success. Performance is only part of the transformation story – there must be some feedback on the culture and value in how we operate?"

*Individual appraisal:* AgileTeam members also work experientially as individuals to appraise Spotify's effects on them. For example, a Scrum Master overseeing a specific agile method did not only appraise the value of the method as proposed by Spotify literature but also its impact on their role and responsibilities. As one Scrum Master (SM) explained:
> "I am also tasked with team evaluations and examine how individuals typically evaluate their team colleagues during the large-scale transformation. This provides me with a point of reference in the hope of getting a more accurate picture of the team and an agile compatible appraisal of team members."

*Reconfiguration:* The appraisal of Spotify by members of AgileTeam had led to numerous top-down reconfiguration attempts to redefine procedures or modify practices. A Senior Technical Manager (STM) within AgileTeam described how:
> "Leaders should be cautious on how software metrics and anecdotal appraisal statements are presented across the organization. Productivity data should not be framed in a competitive sense especially at higher-levels within the organization because teams can strive to undermine or gamify how we carryout evaluations on the large-scale agile transformation."

Managers within FinanceCo regularly revisited their various practices and metrics used to report performance and explore whether restructuring teams or processes could yield better results and reapportion workloads more fairly. However, it was reported by the Senior Vice President (SVP) that "*the frequency of doing so could undermine the transformation process.*"

**CONCLUSION AND FUTURE RESEARCH**

Our research describes the lack of theoretical development on large-scale agile transformations. In this research, we explain how NPT can be applied in research and practice (and a first to do so in the IS field). NPT is applied as a theoretical foundation to investigate the normalization of large-scale agile transformations. This study is the first that goes beyond simple adherence to large-scale agile practices, and actually examines (i) which practices are sustained, and (ii) whether this sustained use actually creates value. In addition, this study also demonstrates that NPT is effective in drawing out key assumptions and uncovering pitfalls associated with the transformation process. In this case study, we uncover key factors which ultimately inhibited the normalization of a large-scale agile transformation and eventually led to the abandonment of the agile method. This research indicates that there is an apparent need to develop an overarching theory which can explain a large-scale agile transformation process. As part of our future research, we will expand on our literature review to map key IS theoretical developments with NPT theoretical constructs. We identified the need to build models that capture the dynamic nature of the transformation process and weave narratives around stakeholders and sequences of events to explain how transformations evolved over time; why they evolved in the way they did; and how they become implemented, embedded, integrated, and evaluated in practice. Therefore, to further ensure rigor and relevance of our NPT theoretical developments and to build a cumulative tradition on IS theories, we will also conduct multiple case studies on large-scale transformations and conduct interviews among multiple stakeholders in global software organizations to better explain the normalization of transformations. We also envisage that the contributions of this research will be far reaching and support new research developments around IS transformation, such as technology transformation and digital transformation.

**ACKNOWLEDGMENTS**

This work was supported with the financial support of the Science Foundation Ireland grant 13/RC/2094 and co-funded under the European Regional Development Fund through the Southern & Eastern Regional Operational Programme to Lero - the Irish Software Research Centre (www.lero.ie).